\documentclass[a4paper]{jpconf}
\usepackage{graphicx}
\usepackage{pgf}
\usepackage[greek,british]{babel}

\newcommand{\TpA}{\ensuremath{\left<T_\mathrm{pA}\right>}}
\newcommand{\Drecoil}{\ensuremath{\Delta_{\rm{recoil}}}}

\begin{document}
\title{Event activity-dependence of jet production in p--Pb collisions
       at $\sqrt{s_{\rm{NN}}} = 5.02$\,TeV measured with semi-inclusive
       hadron+jet correlations by ALICE}

\author{Filip Krizek for the ALICE Collaboration}

\address{Nuclear Physics Institute of CAS, Rez 130, CZ 25068, Czech Republic}

\ead{filip.krizek@cern.ch}

\begin{abstract}
We report measurement of the semi-inclusive distribution
of charged-particle jets recoiling from a high transverse
momentum ($p_{\rm{T}}$) hadron trigger,
for p--Pb collisions at $\sqrt{s_{\rm{NN}}} =5.02$\,TeV, in p--Pb events classified by event activity.
This observable has been measured in pp and Pb--Pb collisions at the LHC,
 providing a new probe to measure quenching.
Jets are reconstructed from charged particle tracks
using anti-$k_{\rm{t}}$ with $R=0.4$ and low IR cutoff
of jet constituents ($p_{\rm{T,track}}>0.15$\,GeV/$c$).
The complex uncorrelated jet background is corrected by a data-driven approach.
Recoil jet distributions are reported for $15<p_{\rm{T,jet}}^{\rm{ch}}<50$\,GeV/$c$.
Events are classified by signal in the ALICE V0A detector,
which measures forward multiplicity, and ZNA, which measures
the number of neutrons at zero degrees.
This self-normalized observable
does not require scaling of reference distributions by \TpA, thereby avoiding
 the need for geometric modeling.
 We compare the trigger-normalized recoil jet yield for p--Pb
collisions with different event activity to measure the effects
of jet quenching in small systems at the LHC.
\end{abstract}

\section{Introduction}
The high energy densities and temperatures achieved
in collisions of heavy nuclei at the LHC and RHIC
generate a Quark-Gluon Plasma (QGP), characterized by collective
flow of soft particles and the quenching of hard jets \cite{Muller:2015jva}.
Recently, flow-like signatures were
observed also in small systems such as p--Pb \cite{Abelev:2013ppb}, and it is natural to ask
 whether jet quenching is also observed in such systems.

Using semi-inclusive hadron+jet correlations,
 the ALICE Collaboration assessed that
in central Pb--Pb collisions at $\sqrt{s_{\rm{NN}}}=2.76$\,TeV,
  charged jets lose energy of about $(8 \pm 2)$\,GeV/$c$
 in a form of charged particles  which are radiated
outside of a resolution parameter $R=0.5$ cone \cite{Adam:2015pbpb}.
The corresponding parton energy loss in a p--Pb collision
is, however, expected to be much smaller than that in Pb--Pb.
First, the average distance traversed by a parton through the hot medium
created in a p--Pb collision will be shorter than in Pb--Pb collisions.
Second, the energy density achieved in p--Pb will be lower.
According to  \cite{Tyw:2014ppb}, energy density affects
 the magnitude of the transport parameter $\hat{q}$ which appears
in the  BDMPS formula \cite{BDMPS:1997qh} for medium-induced parton energy loss.
In p--Pb, $\hat{q}$  should be about seven times smaller than in Pb--Pb \cite{Tyw:2014ppb}.

Experimental searches for medium-induced modification
of inclusive jet production in small systems were carried out by
 PHENIX at RHIC \cite{Adare:2016aa} and ATLAS \cite{Aad:2015ppb} and ALICE \cite{Adam:2015ppb,Adam:2016ppb} at the LHC.
Observed events were classified according to measured event activity (EA),
 i.e., particle multiplicity or total energy detected in a forward
detector which is assumed to be correlated with collision geometry.
PHENIX has shown that inclusive jet
production is not suppressed in minimum bias d--Au collisions at $\sqrt{s_{\rm NN}}=200$\,GeV.
However, for events selected by the measured EA,
jet production is suppressed for large EA (assumed to be ``central'' d--Au collisions)
and enhanced for small EA (assumed to be ``peripheral'').
ATLAS reported similar modification of inclusive jet production as a function of EA
for jets measured in p-going direction in p--Pb collisions
at  $\sqrt{s_{\rm NN}}=5.02$\,TeV.
It is however challenging to directly
correlate measured EA with collision geometry  in small systems, as
various conservation laws and fluctuations play an important role
and induce bias in the geometric modeling \cite{Maj:2014ppb, Abelev:2015cnt}.
This bias can be avoided by constructing observables which allow to identify medium-induced effects
 without the need to know the relation between EA and collision geometry.
This analysis utilizes such an observable, based on the coincidence measurements of
 a high-$p_{\rm{T}}$ semi-inclusive hadron and recoiling jets \cite{Adam:2015pbpb}.

\section{Data Analysis}
The analysis is based on data from p--Pb collisions at  $\sqrt{s_{\rm NN}}=5.02$\,TeV measured by ALICE \cite{Aamodt:2008zz}
in 2013. The per-nucleon momenta of the
two beams in this run were imbalanced,
with the nucleon-nucleon center-of-mass at rapidity $y_{\rm{NN}} = 0.465$ in
the proton beam direction.
        Events were selected online by a Minimum Bias (MB) trigger, consisting of the coincidence of signals
in the V0A and V0C forward scintillator arrays. The offline event selection
rejected events with multiple primary vertices (pile-up events),
and constrained the primary vertex position along the beam axis to $|v_{z}|<10$\,cm.
After all event selection cuts, the number of events in
the analysis is $96 \times 10^{6}$.
EA classification is based upon signals from V0A, with a pseudorapidity
acceptance of $2.8 < \eta < 5.1$,
and ZNA, a neutron calorimeter at zero degrees relative to the beam
direction, at a distance 112.5 m from the vertex diamond
  \cite{Abelev:2015cnt}.  Both detectors are in the Pb-going direction.

The tracking system acceptance covers pseudorapidity
$|\eta| < 0.9$ over the full azimuth, with tracks
reconstructed in the range $0.150 < p_{\rm{T}} < 100$\,GeV/$c$.
The semi-inclusive hadron+jet correlation measurement \cite{Adam:2015pbpb}
 only utilizes events which have a charged, high-$p_{\rm{T}}$ track
(trigger track (TT)).
Since the method considers two exclusive TT transverse momentum bins,
the MB population was divided randomly into two independent subsets.
The first subset was used to search for events where
 TT had transverse momentum in the range
 $12<p_{\rm{T,trig}}<50$\,GeV/$c$ and  the second subset was
used to search for events with TT in the range  $6<p_{\rm{T,trig}}<7$\,GeV/$c$.
These TT selections are labeled as TT\{12,50\} and TT\{6,7\}, respectively.
 If more than one TT candidate was found in a given event, one track was chosen at random.

\begin{figure}[htbp]
\begin{center}
\resizebox{0.525\columnwidth}{!}{\includegraphics{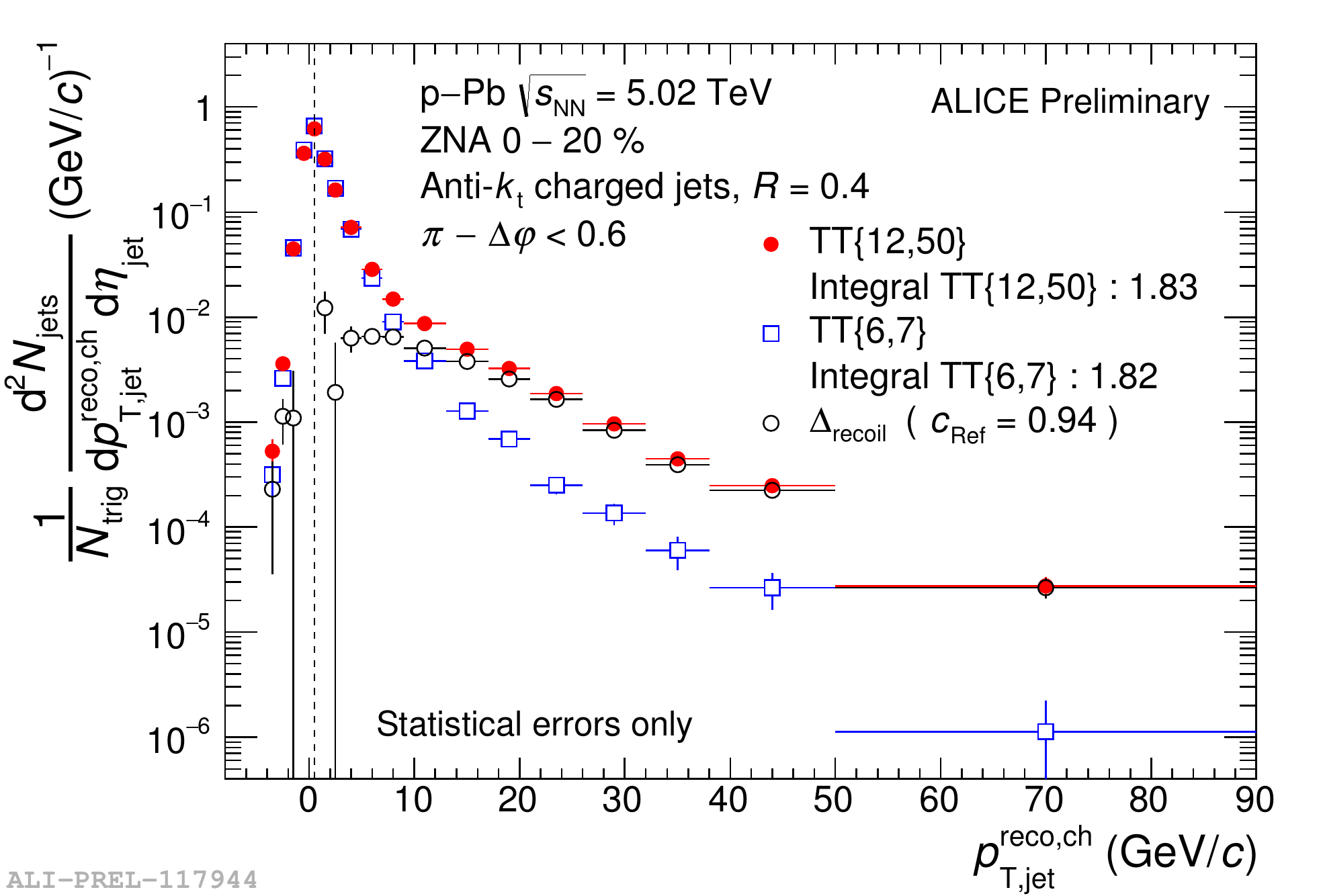}}
\resizebox{0.445\columnwidth}{!}{\includegraphics{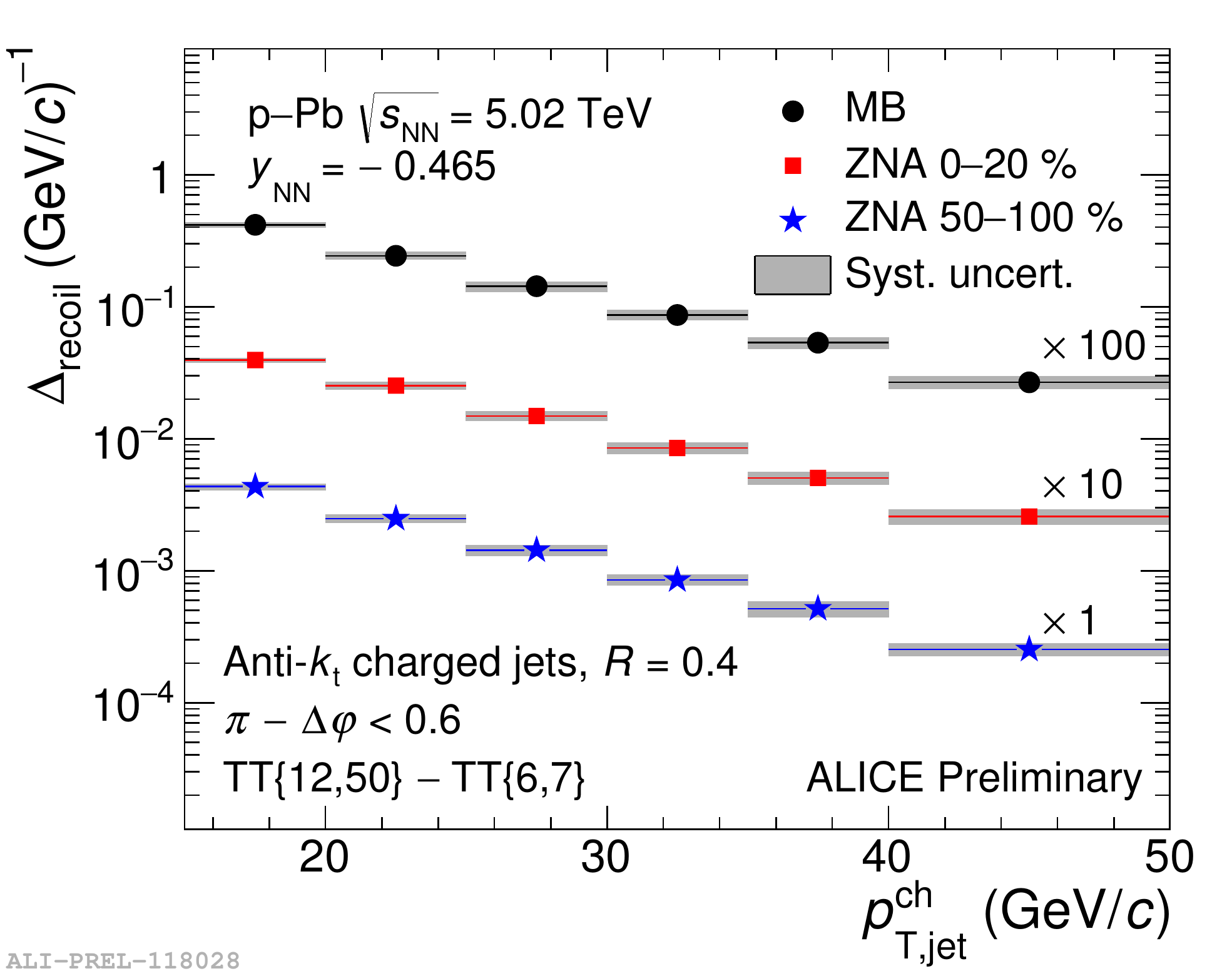}}
\end{center}
 \caption{ Left: Semi-inclusive distributions of charged jets recoiling from
          a high-$p_{\rm{T}}$ hadron trigger in p--Pb collisions
          at $\sqrt{s_{\rm{NN}}} = 5.02$\,TeV, for EA in the ZNA 0--20\% bin.
          Right: Fully corrected $\Delta_{\rm{recoil}}$ spectra for different EA biases.
          See text for details.}
\label{fig:DeltaRecoilSpectra}
\end{figure}

Jets were reconstructed from charged tracks
using the anti-$k_{\rm{t}}$ algorithm \cite{FastJetAntikt}
with resolution parameter $R=0.4$ and
the boost-invariant $p_{\rm{T}}$ recombination scheme.
Jet candidates were accepted for further analysis if their centroid and area satisfied
$|\eta_{\rm{jet}}|<0.9-R$ and $A_{\rm{jet}}>0.6\,\pi\,R^{2}$, respectively.
Reconstructed jet momenta $p_{\rm{T,jet}}^{\rm{raw,ch}}$
are corrected for the mean underlying event density $\rho$
which is assessed on event by event basis
using the standard area based approach \cite{FastJetPileup},
$p_{\rm{T,jet}}^{\rm{reco,ch}} = p_{\rm{T,jet}}^{\rm{raw,ch}} - \rho\,A_{\rm{jet}}\,.$
Once a TT was found, it was correlated with jets that
are oriented nearly back-to-back in azimuth
relative to the TT.
 The azimuthal angle contained between the TT and the jet
was required to be larger than $\pi - 0.6$. Fig.~\ref{fig:DeltaRecoilSpectra}, left panel, shows an example
of the raw per-trigger normalized yield of recoil jets
associated to TT\{12,50\} and TT\{6,7\}, measured as a function
of jet transverse momentum. The region of negative and small positive
 $p_{\rm{T,jet}}^{\rm{reco,ch}}$ is dominated by uncorrelated
jet yield from the underlying event,
 while the region for large positive  $p_{\rm{T,jet}}^{\rm{reco,ch}}$
is dominated by recoil jet yield that is correlated with TT.

The trigger-normalized semi-inclusive jet recoil distribution in this
analysis is equivalent to measuring the ratio of two cross
sections, with numerator the coincidence cross section for
a trigger hadron and jet in the acceptance, and denominator the inclusive cross
section for a hadron in the trigger $p_\mathrm{T}$-interval~\cite{Adam:2015pbpb}. The
observable is calculable using NLO pQCD \cite{deFlorian:2009fw,Adam:2015pbpb}. Jet quenching measurements require comparison to a reference spectrum in which quenching is not expected to occur. Since inclusive cross sections in different systems are related by the nuclear overlap integral \TpA, the reference distribution for this observable scales as

\begin{equation}
\label{eq:Canc}
 \frac{1}{\sigma_\mathrm{ref}^{\mathrm{pA}\rightarrow \mathrm{h}+X }}\frac{{\mathrm d}^{2}\sigma_\mathrm{ref}^{\mathrm{pA}\rightarrow \mathrm{h+jet}+X}}{{\mathrm d}p_{\mathrm{T,jet}}^{\mathrm{ch}}{\mathrm d}\eta_{\mathrm{jet}}}\bigg|_{p_{\rm{T,h}}\in \mathrm{TT}} =
 \frac{1}{\TpA \sigma^{\mathrm{pp}\rightarrow \mathrm{h}+X }}\frac{\TpA {\mathrm d}^{2}\sigma^{\mathrm{pp}\rightarrow \mathrm{h+jet}+X}}{{\mathrm d}p_{\mathrm{T,jet}}^{\mathrm{ch}}{\mathrm d}\eta_{\mathrm{jet}}}\bigg|_{p_{\rm{T,h}}\in \mathrm{TT}}\,.
\end{equation}

\noindent
The observable is self-normalized, and the factor \TpA\ in numerator
and denominator cancel identically. This observable therefore provides a
measurement of jet quenching with no dependence on \TpA, and consequently no
requirement to interpret EA in terms of collision geometry, thereby avoiding the
largest systematic uncertainty of jet quenching measurements in small systems
using inclusive processes.

The raw recoil jet yield includes jets that are uncorrelated to the trigger hadron, including jets generated in multi-partonic interactions. The uncorrelated jet yield is removed by measuring the distribution \Drecoil, which is the difference between the recoil jet distributions in the two TT classes~\cite{Adam:2015pbpb},
\begin{equation}
\label{eq:Dr}
 \Drecoil = \frac{1}{N_\mathrm{trig}} 
         \frac{{\mathrm d}^{2}N_\mathrm{jet}}{{\mathrm d}p_{\mathrm{T,jet}}^{\mathrm{ch}}{\mathrm d}\eta_{\mathrm{jet}}}\bigg|_{\mathrm{TT}\{12,50\}} -  c_{\mathrm{Ref}} \cdot 
         \frac{1}{N_\mathrm{trig}}
        \frac{{\mathrm d}^{2}N_\mathrm{jet}}{{\mathrm d}p_{\mathrm{T,jet}}^{\mathrm{ch}}{\mathrm d}\eta_{\mathrm{jet}}}\bigg|_{\mathrm{TT}\{6,7\}}\,\,,
\end{equation}
\noindent
where the factor $c_{\mathrm{Ref}}$ accounts for invariance of the jet density
with TT-class~\cite{Adam:2015pbpb}, determined in this analysis from the yields in $0<p_{\rm{T,jet}}^{\rm{reco,ch}}<1$\,GeV/$c$, and with value between 0.92 and 0.99. We note that \Drecoil\ is a differential, not absolute, observable. It measures the change in recoil jet yield as the trigger hadron $p_\mathrm{T}$ is increased from the lower  to higher TT-interval~\cite{Adam:2015pbpb}. This approach enables the measurement of reconstructed jets over a wide range in jet $p_\mathrm{T}$ and $R$, in the presence of large backgrounds~\cite{Adam:2015pbpb}.

The raw $\Delta_{\rm{recoil}}$ distribution was corrected via unfolding for momentum smearing due to
instrumental effects and local background fluctuations.
The fully corrected  $\Delta_{\rm{recoil}}$ distributions
for different EA selections are shown in the right panel
in Fig.~\ref{fig:DeltaRecoilSpectra}. The main source of systematic uncertainty
is the track reconstruction efficiency, resulting in 4--10\,\% uncertainty
on the spectra. Other sources of uncertainty are unfolding, choice of $c_{\rm{Ref}}$,
$\rho$ estimator, track momentum smearing etc.
They are typically below 4\,\%.
The cumulative systematic uncertainty is the quadratic sum of all contributions.

\begin{figure}[htbp]
\begin{center}
\resizebox{0.39\columnwidth}{!}{\includegraphics{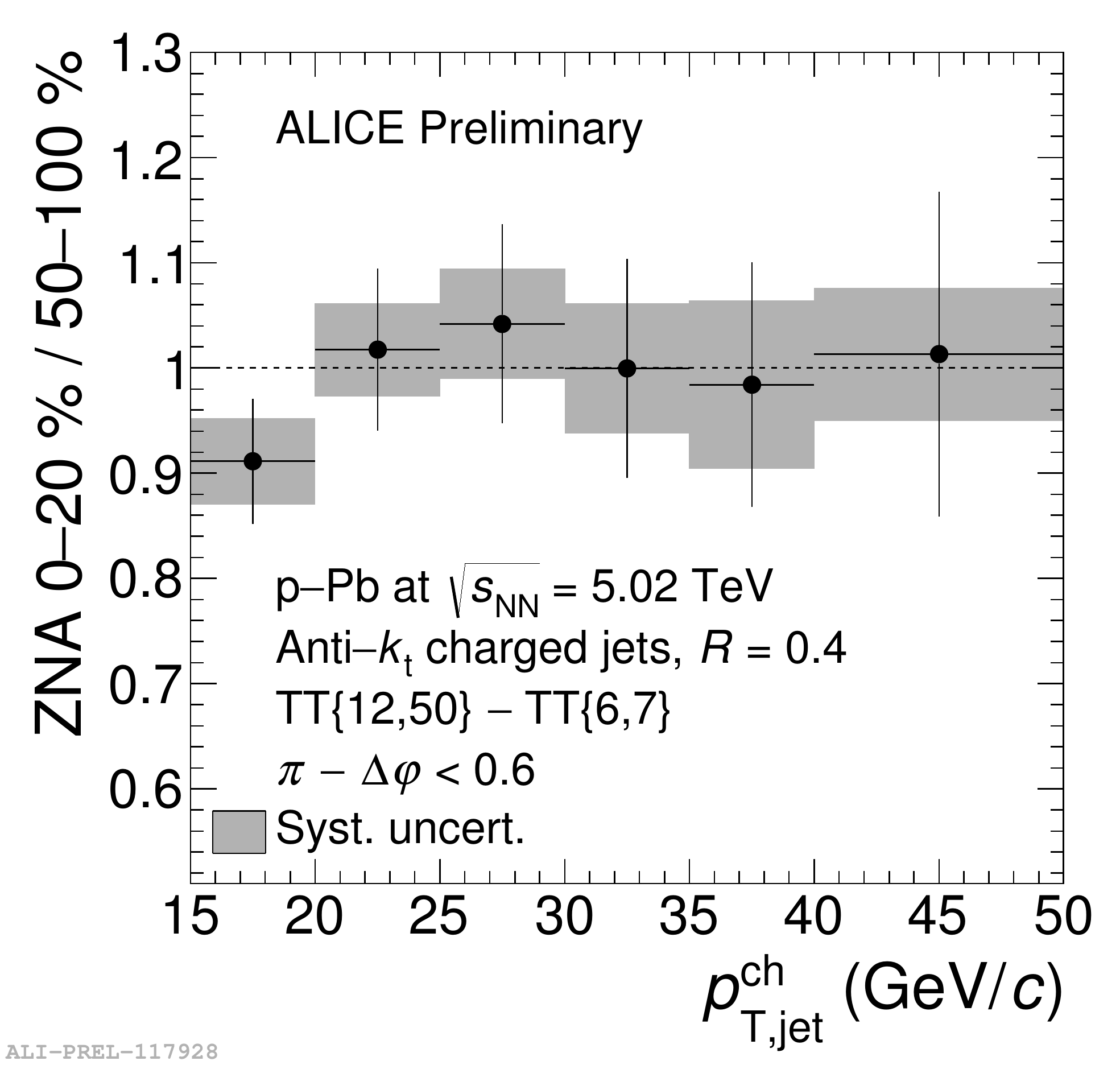}}
\resizebox{0.39\columnwidth}{!}{\includegraphics{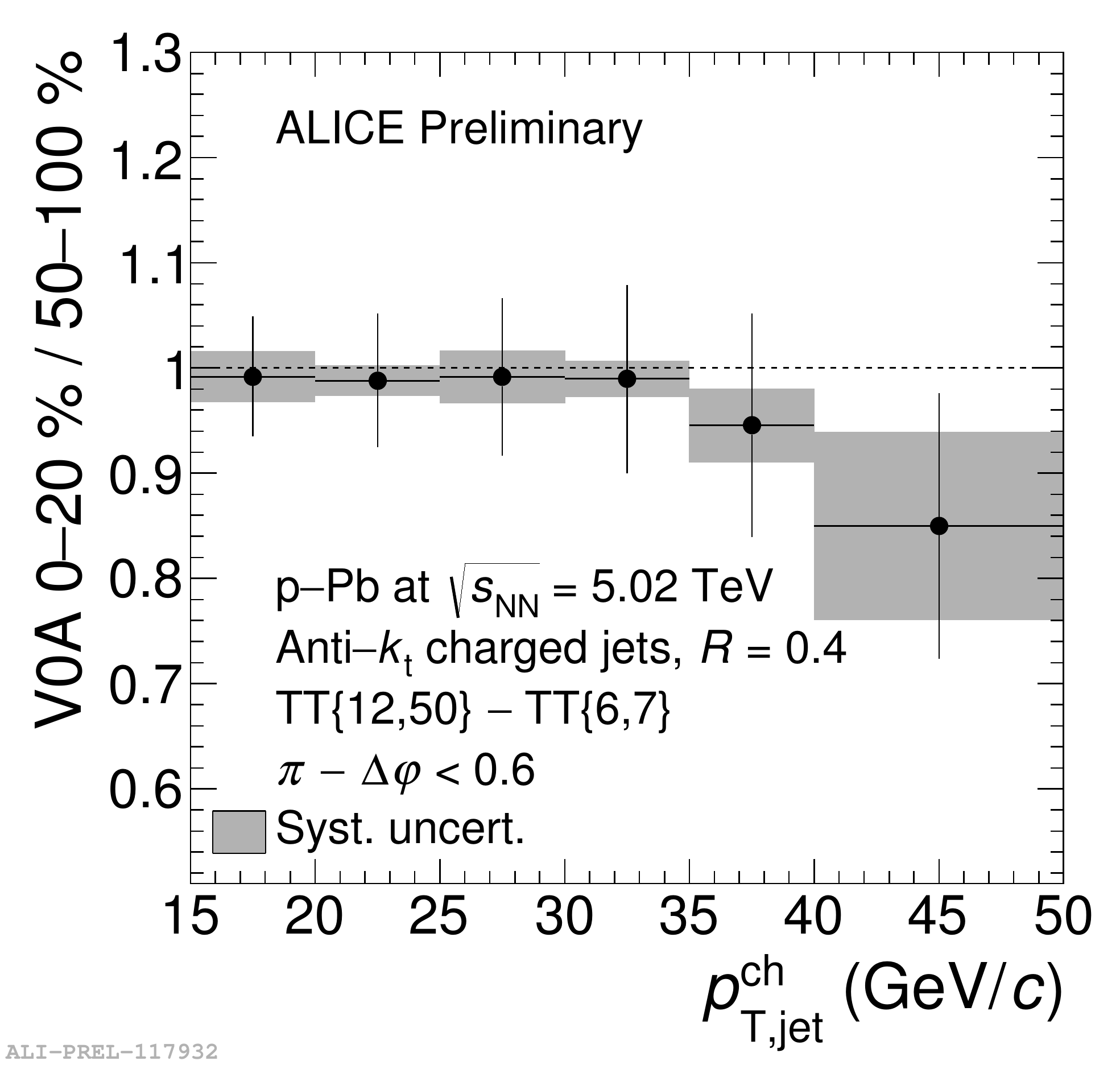}}
\end{center}
 \caption{Ratio of $\Delta_{\rm{recoil}}$ distributions for events
with high and low EA measured in p--Pb collisions at $\sqrt{s_{\rm{NN}}} =5.02$\,TeV.
Left panel: V0A 0--20\,\% / 50--100\,\%; right panel: ZNA 0--20\,\% / 50--100\,\%.
The grey boxes show the systematic uncertainty of the ratio, which accounts for the
correlated uncertainty of numerator and denominator.}
\label{fig:DeltaRCP}
\end{figure}

Jet quenching may result in transport of jet energy out of the jet cone, which would suppress
$\Delta_{\rm{recoil}}$ at fixed $p_{\rm{T,jet}}^{\rm{ch}}$.
If jet quenching is more likely to occur in events with greater EA,
the $\Delta_{\rm{recoil}}$ spectrum ratios
shown in Fig.~\ref{fig:DeltaRCP} will be suppressed below unity.
However, the ratio is consistent with unity at all $p_{\rm{T,jet}}^{\rm{ch}}$
within the statistical error and the systematic uncertainty in all panels, indicating
negligible jet quenching effects relative to the uncertainties.
These data can nevertheless provide a limit  on the magnitude of medium-induced energy transport
to large angles. For this estimate, we assume that the average magnitude of energy transported
out-of-cone is independent of $p_{\rm{T,jet}}^{\rm{ch}}$.
We parameterize the 50--100\,\%  and 0--20\,\% $\Delta_{\rm{recoil}}$ distributions with the
exponential functions
$\Delta_\mathrm{recoil}|_{50-100\%} = a \exp\left[- p_{\mathrm{T,jet}}^{\mathrm{ch}} / b \right]$
and $\Delta_\mathrm{recoil}|_{0-20\%} = 
 a \exp\left[- \left(p_{\mathrm{T,jet}}^{\mathrm{ch}} + \bar{s} \right)/ b \right]$,
 with common fit parameters $a$ and $b$. The ratios in Fig.~\ref{fig:DeltaRCP} are then
expressed in terms of an average shift $\bar{s}$
in $p_{\rm{T,jet}}^{\rm{ch}}$ between low and high EA events,
$ \Delta_\mathrm{recoil}|_{0-20\%}\,/\,\Delta_\mathrm{recoil}|_{50-100\%} = \exp\left(-\bar{s}/b\right)$.
Fits to $\Delta_{\rm{recoil}}$ for $R = 0.4$ over
the range $15 < p_{\rm{T,jet}}^{\rm{ch}} < 50$\,GeV/$c$ give $b = 9.43 \pm 0.31$\,GeV/$c$ for ZNA
50--100\,\% and  $b = 9.76 \pm 0.29$\,GeV/$c$ for V0A 50--100\,\%.
Fits to the ratios in Fig.~\ref{fig:DeltaRCP} then yield
$\bar{s}= \left(0.22 \pm 0.35_{\mathrm{stat}} \pm 0.05_{\mathrm{syst}}\right)$\,GeV/$c$
for ZNA 0--20\,\% and
$\bar{s}= \left(0.22 \pm 0.31_{\mathrm{stat}} \pm 0.05_{\mathrm{syst}}\right)$\,GeV/$c$
for V0A 0--20\,\%, consistent with zero.
These values should be compared with the significant shift
$\bar{s}= \left(8 \pm 2_{\mathrm{stat}}\right)$\,GeV/$c$
measured in central Pb--Pb collisions at $\sqrt{s_{\rm{NN}}}=2.76$\,TeV,
corresponding to finite energy loss \cite{Adam:2015pbpb}. The measurement of $\bar{s}$ also provides a limit on
 out-of-cone energy transport due to jet quenching in p--Pb collisions. Under
the assumption that its magnitude is independent of
$p_{\rm{T,jet}}^{\rm{ch}}$, the medium-induced out-of-cone energy transport for
events
 with high relative to low EA is less than 0.7\,GeV/$c$, at 90\% confidence,
for jets with $R = 0.4$ and $15 < p_{\rm{T,jet}}^{\rm{ch}}< 50$\,GeV/$c$.

{\bf{Acknowledgement:}} The work has been supported by the grant LG15052 of the Ministry of Education Youth and Sports of the Czech Republic.

\end{document}